\documentclass[amsmath,amssymb,showpacs,floatfix,twocolumn,pra,superscriptaddress]{revtex4}
\usepackage{graphicx}
\usepackage{dcolumn}
\usepackage{bm}

\def\beq{\begin{equation}}
\def\eeq{\end{equation}}
\def\a{\alpha}
\def\b{\beta}
\def\g{\gamma}
\def\o{\omega}

\newcommand{\bL}{\mathbf{L}}
\newcommand{\bk}{\mathbf{k}}

\newcommand{\bq}{\mathbf{q}}
\newcommand{\br}{\mathbf{r}}
\newcommand{\bR}{\mathbf{R}}

\newcommand{\bd}{\mathbf{d}}

\newcommand{\bE}{\mathbf{E}}

\newcommand{\la}{\langle}
\newcommand{\ra}{\rangle}

\newcommand{\eq}[1]{(\ref{eq:#1})}
\newcommand{\eqname}[1]{\label{eq:#1}}

\begin{document}

\title{Probing quasi-particle states in strongly interacting atomic gases
by momentum-resolved Raman photoemission spectroscopy}

\author{Tung-Lam Dao}
\affiliation{Centre de Physique Th{\'e}orique, Ecole Polytechnique, CNRS,
91128 Palaiseau Cedex, France}
\author{Iacopo Carusotto}
\affiliation{CNR-INFM BEC Center and universit\`a di Trento, 38050
Povo, Italy} \affiliation{Institute for Quantum Electronics, ETH
Z\"urich, 8093 Z\"urich, Switzerland}

\author{Antoine Georges}
\affiliation{Centre de Physique Th{\'e}orique, Ecole Polytechnique, CNRS,
91128 Palaiseau Cedex, France}
\affiliation{Coll\`ege de France, 11 place Marcelin Berthelot, 75005 Paris, France}

\begin{abstract}
We investigate a momentum-resolved Raman spectroscopy technique
which is able to probe the one-body spectral function and the
quasi-particle states of a gas of strongly interacting ultracold
atoms. This technique is inspired by Angle-Resolved Photo-Emission
Spectroscopy, a powerful experimental probe of electronic states
in solid-state systems. Quantitative examples of experimentally
accessible spectra are given for the most significant regimes
along the BEC-BCS crossover. When the theory is specialized to RF
spectroscopy, agreement is found with recent experimental data.
The main advantages of this Raman spectroscopy over existing
techniques are pointed out.
\end{abstract}

\date{\today}
 \maketitle

\section{Introduction}
The recent advances in the cooling, trapping, and manipulation of
ultra-cold atomic gases using mostly optical beams have given
birth to the new field of  ``condensed matter physics with light
and atoms''. Key issues of the physics of strongly correlated
quantum systems can now be addressed in this new context from a
completely different perspective. Remarkable milestones in this
respect have been the observation of the superfluid to
Mott-insulator transition in a system of bosons in an optical
lattice~\cite{greiner_mott_nature_2002}, the direct imaging of the
Fermi surface in a degenerate Fermi
gas~\cite{kohl_fermisurface_prl_2005}, the demonstration of
superfluidity in an interacting Fermi
gas~\cite{Superfluid,Jochim,Zwierlein,Kinast,Bourdel}, and the
recent
observation~\cite{Jordens_mott_nature_2008,Schneider_mott_science_2008}
of an incompressible Mott insulating regime of fermionic atoms
\cite{Lorenzo,Scarola} trapped in the periodic potential of
optical lattices.

Central objects in the theoretical and experimental study of
quantum many-body systems are the low-energy excited states, and
the possibility of describing those states in terms of
quasi-particle excitations having a long lifetime and a
well-defined dispersion relation for the excitation energy as a
function of momentum~\cite{mbt_books,mahan_book}. There is in fact
abundant experimental and theoretical evidence that quasi-particle
states can be highly unconventional in strongly correlated
systems, and significantly depart from the ones predicted by
Landau Fermi-liquid
theory~\cite{damascelli_rmp_2003,norman_pheno_prb_1998}. An
understanding of quasi-particle states is therefore an essential
step in the direction of building a complete description of the
peculiar electronic, magnetic, and optical properties that have
been observed in a variety of strongly correlated materials.

A commonly used probe of the electronic states in solid materials
is the so-called angle-resolved photoemission spectroscopy
(ARPES)~\cite{damascelli_ARPESintro_physscripta_2004}. This
technique consists in measuring the energy and momentum
distribution of the electrons that are emitted from the solid when
it is exposed to a beam of energetic photons. In the simplest
approximation, the distribution of the emitted electrons is in
fact proportional to the one-body spectral function of the
electrons in the solid.

Inspired by the success of ARPES in solid-state systems, we
recently proposed momentum-resolved stimulated Raman spectroscopy
as a probe of quasi-particle states in strongly correlated atomic
Fermi gases~\cite{DaoPRL,TLDaoThesis}. In a stimulated Raman
process, atoms are transferred from the gas into a different
internal state by absorbing a photon from a laser beam and
immediately reemitting it into another beam of different frequency
and wavevector. For a given intensity and duration of the Raman
pulses, the momentum distribution of extracted atoms is measured
as a function of the wavevector and frequency of the Raman beams:
analogously to electronic ARPES, this distribution is proportional
to the one-body spectral function of the atoms in the gas.
Alternative schemes that use Raman processes to probe the
microscopic properties of atomic gases have been proposed
in~\cite{Japha,Luxat,Raman2,Carusotto,Blakie,Duan,Yi}.
Experimental applications of the related Bragg scattering
technique have been reported
in~\cite{Bragg,Stamper-Kurn,Stoferle,Veeravali}.

This momentum-resolved stimulated Raman technique is a powerful
extension of the RF spectroscopy technique that has been recently
used to study strongly interacting atomic Fermi
gases~\cite{Zoller,Grimm,Ketterle_Varenna_2008}. A pioneering
experimental demonstration of the application of momentum-resolved
RF spectroscopy to a strongly correlated atomic gas has recently
appeared~\cite{Jin:RFspectroscopy}: the energy dispersion of the
occupied quasi-particle states has been measured at several points
along the BEC-BCS crossover and has shown clear evidence of a
pairing gap. Momentum-resolved RF spectroscopy in atomic Fermi
gases has also been discussed recently in
Ref.~\cite{Chen_condmat_2008}.

In this paper we extend the discussion of Ref.~\cite{DaoPRL} and
we give a comprehensive discussion of the promise of
momentum-resolved stimulated Raman spectroscopy to investigate the
microscopic properties of strongly interacting Fermi gases. On one
hand, RF spectroscopy naturally arises as a special case of Raman
spectroscopy with a vanishing transferred momentum: the
theoretically calculated spectra are in good agreement with the
experimental observations of Ref.\onlinecite{Jin:RFspectroscopy}.
On the other hand, Raman spectroscopy offers several significant
advantages as compared to the RF one,  e.g. spatial selectivity to
eliminate inhomogeneous broadening due to the trap potential,
tunability of the transferred momentum from below to well above
the Fermi momentum, weaker sensitivity to final-state
interactions.

The paper is organized as follows: in Sec.\ref{sec:general}, we
review the general theory of the Raman spectroscopy and we provide
explicit formulas for the observable signal in terms of the
one-body spectral function of the gas. In Sec.\ref{sec:nonint},
the advantage of the Raman spectroscopy over the RF one are
illustrated on the simplest case of an ideal Fermi gas. In
Sec.\ref{sec:strongly}, we show how the general theory is able to
reproduce the experimental observations of
Ref.\onlinecite{Jin:RFspectroscopy} for a strongly interacting
Fermi gas along the BEC-BCS crossover. The promise of an inverse
Raman spectroscopy to investigate the dispersion of the empty
quasi-particle states is discussed in Sec.\ref{sec:inverse}.
Concluding remarks are given in Sec.\ref{sec:conclu}.

\section{General theory of the Raman photo-emission technique}
\label{sec:general}

\subsection{The optical process}

\begin{figure}[!ht]
  \includegraphics[width=4cm,clip=true]{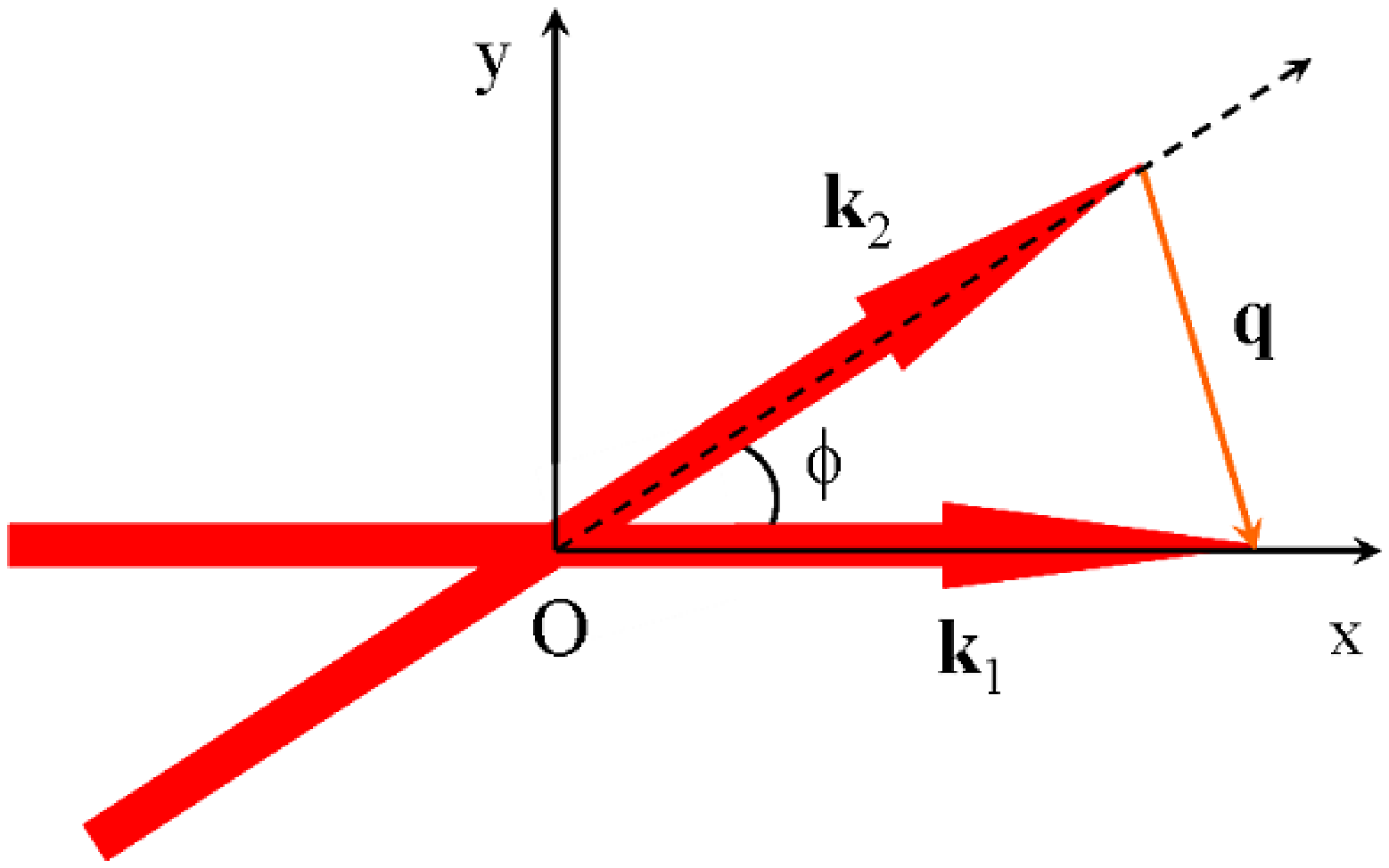}\\
  \includegraphics[width=5cm,clip=true]{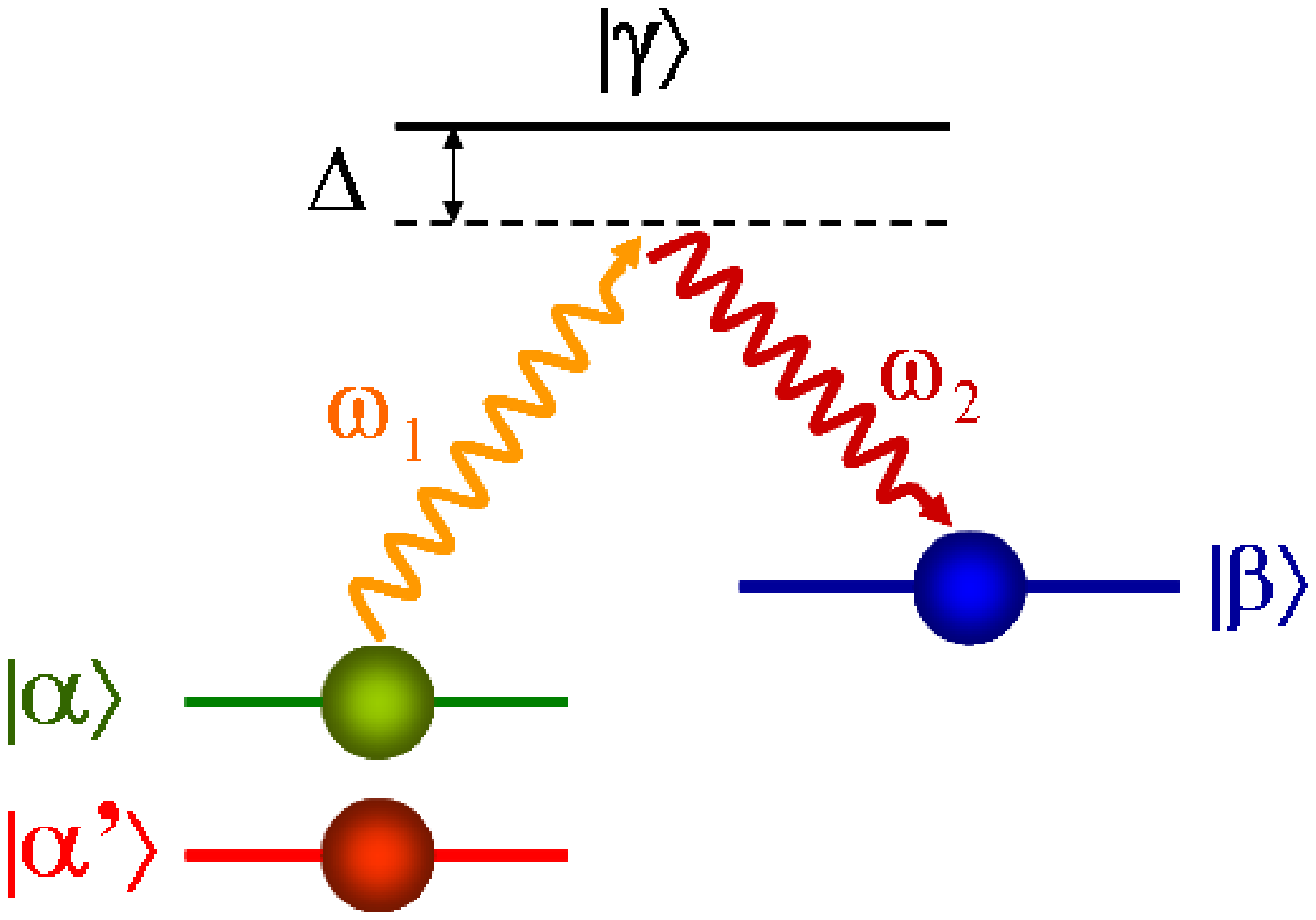}
  \caption{Upper panel: geometrical configuration of the Raman laser beams. Lower panel:
scheme of the atomic levels involved in the Raman
process.\label{Raman}}
\end{figure}

In this section, we review the general theory of the stimulated
Raman spectroscopy technique to probe the one-particle excitations
of a many-atom gas that was originally proposed in~\cite{DaoPRL}.
Throughout the whole paper we will concentrate our attention on
the case of a two-component mixture of fermionic atoms in two
internal states $\a$ and $\a'$, but the extension to a generic
system of fermions or bosons is straightforward. Generally, the
$\a$ and $\a'$ states are long-lived hyperfine components of the
electronic ground state of the atom. Atoms are transferred from
the $\a$ state to a third hyperfine component  $\b$ by a pair of
laser beams in the Raman configuration schematically shown in
Fig.\ref{Raman}. The laser fields are assumed to be far from
resonance with the intermediate excited state $\gamma$ so that
spontaneous emission events can be neglected.

The atom-laser interaction can be described by the standard dipolar Hamiltonian
\begin{equation}
V_{\textrm{dip}}=-\bd \cdot \left [ \bE_1(\br,t) + \bE_2(\br,t) \right].
\end{equation}
where the laser fields are assumed to be classical and have the form
\begin{equation}
\bE_{1,2}(\br,t)=\tilde{\bE}_{1,2}(\br)\,
e^{i(\bk_{1,2}\br-\omega_{1,2}t)}+\textrm{c.c.}
\end{equation}
The envelopes $\tilde{\bE}_{1,2}(\br)$ are slowly varying on the
scale of the carrier wavevectors $\bk_{1,2}$ and account for the
transverse profile of the laser beams. The corresponding Rabi
frequencies are defined in terms of the electric dipole matrix
elements of the optical transitions connecting the $\alpha,\beta$
states to the common excited state $\gamma$ as
$\hbar\Omega_{1}(\br)=\bd_{\gamma\a}\tilde{\bE}_{1}(\br)$ and
$\hbar\Omega_{2}(\br)=\bd_{\gamma\b}\tilde{\bE}_{2}(\br)$.

Once we perform the rotating wave approximation and we eliminate
the intermediate excited state $\gamma$ under the condition
$\Omega_{1,2},\,\Gamma_\gamma \ll \Delta$ ($\Gamma_\gamma$ is here
the natural linewidth of the excited $\gamma$ state), the
light-matter interaction can be reduced to the simple form:
\begin{equation}
\hat{V}(t)=\hat{V}\,e^{-i\omega t} + \textrm{h.c.}\eqname{Raman}
\end{equation}
in which $\hat{V}=\int d\br\, \Omega_{e}(\br,t)\,
\hat{\psi}^{\dagger}_{\b\br}\, \hat{\psi}_{\a\br}\,e^{i\bq.\br}$.
The basic process described by the Hamiltonian \eq{Raman} consists
of the transfer of atoms from the $\a$ to the $\g$ state by
conservative Raman processes.
$\Omega_{e}(\br,t)=\Omega_{1}(\br,t)\,\Omega^{*}_{2}(\br,t)/\Delta$
is the Rabi frequency of the effective Raman coupling,
$\bq=\bk_{1}-\bk_{2}$ is the transferred momentum, and
$\omega=\omega_{1}-\omega_{2}$ is the transferred energy.

Note that RF spectroscopy with a single electromagnetic field at
frequency $\omega$ is described by a Hamiltonian of exactly the
same form \eq{Raman} with a Rabi frequency
$\Omega_e(\br)=\bd_{\b\a}\tilde{\bE}(\br)$ and a vanishing
transferred momentum $\bq=0$. All the discussion that follows is
then straighforwardly extended to the case of RF spectroscopy by
simply setting $\bq=0$.

\subsection{The Raman emission rate}

We first calculate the Raman transfer rate in the simplest case in
which atoms in the final state $\beta$ do not interact with the
atoms left behind in the $\a,\a'$ states and furthermore do not
feel any trapping potential. Under these assumptions, their
momentum is conserved while they freely propagate in space and one
can envisage to perform a momentum-resolved measurement along the
lines of the RF experiment of~\cite{Jin:RFspectroscopy}.

In terms of the many-body states of the gas, the Raman process
then consists in the excitation of the initial $N$ body state
$|\phi^{N}_{i}\ra$ to an excited $N-1$ body state
$|\phi^{N-1}_{f}\ra$ and the simultaneous out-coupling of one atom
into the $\beta$ state with momentum $\bk$. The rate
$R_\bq(\bk,\omega)$ for this process depends on the wavevector
$\bq=\bk_1-\bk_2$ and frequency $\omega=\o_1-\o_2$ that are
transferred by the Raman beams to the atoms. The total Raman
transfer rate $R^T_\bq(\omega)$ is then obtained by integrating
over final momenta $\bk$ as follows:
\begin{equation}
R^T_\bq(\omega)=\int\!d^3\bk\,R_\bq(\bk,\omega).
\eqname{total}
\end{equation}
Direct application of the Fermi golden rule in the grand-canonical ensemble leads to
\begin{multline}
R_\bq(\bk,\omega)=\frac{2\pi}{\hbar}\, \sum_{i,f} \frac{e^{-\bar{E}_i/k_B T} }{\mathcal{Z}}\, | M_\bk^{fi} |^2\\
\times \delta(E_f+  \varepsilon_{\bk\b} -E_i -\hbar\omega).
\label{eq:Raman_rate}
\end{multline}
The initial state is assumed to be at thermal equilibrium at a
temperature $T$ with a chemical potential $\mu$;
$\mathcal{Z}=\sum_i \exp(-\bar{E}_i/k_B T)$ is the corresponding
Grand-Canonical partition function. The sums over $i$ and $f$
refers to all the many-body states of the system. The energy
$\bar{E_i}=E_i-\mu N_i$ is rescaled by the number of particles.
The matrix element $M_{\bk}^{fi}$ has the form:
\begin{align}
M_{\bk}^{fi}&=\int d^3\br \,
 \la \bk\b | \psi^{\dagger}_{\b\br} |0\ra \la
\phi_{f}|\psi_{\a\br}|\phi_{i} \ra \, \Omega_{e}(\br)\,e^{i\bq \br}\nonumber\\
&=\int d^3\br \, \la \phi_{f}|\psi_{\a\br}|\phi_{i} \ra \,
\Omega_{e}(\br)\,e^{i(\bq-\bk)\br}
\end{align}
Using the definition of the real space spectral function at finite
temperature ~\cite{mahan_book},
\begin{multline}
A(\br,\br';\omega')=\sum_{i,f} \frac{e^{-\bar{E}_i/k_B T}+e^{-\bar{E}_f/k_B T}}{\mathcal{Z}}\\
\times \la\phi_{i}|\psi^{\dagger}_{\a\br'}|\phi_{f}\ra
  \la\phi_{f}|\psi_{\a\br}|\phi_{i}\ra
\delta(\hbar\omega'+\bar{E}_f-\bar{E}_i)
\eqname{realSpectralFunction}
\end{multline}
the Raman transfer rate can be rewritten as
\begin{multline}
R_{\bq}(\bk,\omega)= \frac{2\pi}{\hbar}
\int\! d^3\br\, \int\! d^3\br'\,\Omega_{e}(\br)\, \Omega^{*}_{e}(\br')\,
e^{i(\bq-\bk)(\br-\br')}\\
\times n_{F}(\varepsilon_{\bk\beta}-\hbar\omega-\mu)\, A(\br,\br';\varepsilon_{\bk\beta}-\hbar\omega-\mu).
\label{eq:RamanRateA}
\end{multline}
Here, the Fermi factor has the usual definition
$n_F(E)=1/[1+\exp(E/k_BT)]$.

\subsection{Spatially homogeneous geometry}
\label{sec:homogeneous}

In the simplest case of a spatially homogeneous system with no
trapping potential and spatially uniform Raman beams
$\Omega_e(\br)=\Omega_e$, the Raman rate per unit volume can be
rewritten in a simplified form:
\begin{multline}
\frac{dR_{\bq}(\bk,\omega)}{d{V}}=\frac{2\pi}{\hbar}\,|\Omega_e|^2\,n_F(\varepsilon_{\bk\beta}-\hbar\omega-\mu)\\
\times A(\bk-\bq,\varepsilon_{\bk\beta}-\hbar\omega-\mu).
\eqname{RamanRateAk}
\end{multline}
that only involves the momentum space spectral function
\begin{multline}
A(\bk,\omega)=\sum_{i,f} \frac{e^{-\bar{E}_i/k_B T}+e^{-\bar{E}_f/k_B T}}{\mathcal{Z}}\\
\times
\left|\la\phi_{f}|\psi_{\a\bk}|\phi_{i}\ra\right|^2\,\delta(\hbar\omega+\bar{E}_f-\bar{E}_i).
\eqname{spectral_homo}
\end{multline}
In the simplest case of a non-interacting gas at zero temperature
$T=0$, the spectral function \eq{spectral_homo} has the form
\begin{equation}
A(\bk,\omega)=\delta(\hbar\omega+\mu-\varepsilon_{\bk\alpha})
\end{equation}

The energy and momentum conservation condition that underlie
\eq{RamanRateAk} have a simple physical interpretation. Initially,
the many-body system is in the state $i$ of energy $E_i^N$ of
momentum $\bk^N_i$; the laser beams consist of $N_{1,2}$ photons
in the two beams of respectively frequencies $\o_{1,2}$ and
momenta $\bk_{1,2}$. At the end of the Raman process, the
many-body system has an energy $E_{f}^{N-1}$ and a momentum
$\bk_f^{N-1}$, while the laser beams contain respectively $N_1-1$
and $N_2+1$ photons. By energy and momentum conservation, one
therefore has that
\begin{eqnarray}
E_i^N&+&N_1\hbar\omega_1+N_2\hbar\omega_2=E_f^{N-1}+(N_1-1)\hbar\omega_1+ \nonumber \\
&+&(N_2+1)\hbar\omega_2+\varepsilon_{\bk\beta} \\
\bk^N_i&+&N_1\bk_1+N_2\bk_2=\bk^{N-1}_f+(N_1-1)\bk_1+ \nonumber
\\ &+&(N_2+1)\bk_2+\bk.
\end{eqnarray}
which reduces to:
\begin{eqnarray}
E_i^N-E_f^{N-1}&=&\varepsilon_{\bk\beta}-\hbar\omega \nonumber \\
\bk_i^N-\bk_f^{N-1}&=&\bk-\bq
\end{eqnarray}
As expected, these conditions perfectly correspond to the frequency and momentum values
at which the spectral function in \eq{RamanRateAk} is evaluated.

\subsection{Local density approximation}
\label{sec:LDA}

In view of concrete applications, it is useful to derive
approximate formulas that can accurately describe the case of
spatially selective Raman processes in trapped systems.  Spatial
selectivity is in fact a key advantage of Raman techniques over RF
spectroscopy, as it allows to avoid inhomogeneous broadening
effects by restricting the Raman out-coupling process to a limited
region of space where the system can be seen as almost uniform.
This possibility is even more important if the system presents
several different phases with macroscopically different
properties.

The simplest way to include the trapping potential $V_\alpha(\bR)$
acting on the $\alpha$ atoms and/or of the spatial profile of the
Raman beams is to perform the so-called Local Density
Approximation (LDA). This approximation is generally accurate as
long as the characteristic length on which the properties of the
system vary is much larger than all microscopic scales of the
system.

Under this approximation, the Raman rate can be written as an
integral over different regions of space,
\begin{multline}
R_\bq(\bk,\omega)=\frac{2\pi}{\hbar}\int \! d^3\bR\,|\Omega_e|^2\,n_F(\varepsilon_{\bk\beta}-\hbar\omega-\mu) \\
\times A(\bk-\bq,\varepsilon_{\bk\beta}-\hbar\omega-\mu;\mu_\bR)
\eqname{LDA1}
\end{multline}
the contribution of each spatial region is approximated by the
prediction \eq{RamanRateAk} for a homogeneous system with an
effective chemical potential $\mu_\bR=\mu-V_\alpha(\bR)$ that
includes the effect of the trap. Here,
$A(\bk,\hbar\omega;\mu_\bR)$ is the spectral function
\eq{spectral_homo} with the local chemical potential $\mu_\bR$.

By pushing further the LDA approximation, one can obtain a simple
formula for the Raman rate also in the presence of an external
potential $V_\beta(\bR)$ acting on the atoms in the $\beta$ state,
\begin{multline}
R_\bq(\bk,\omega)=\frac{2\pi}{\hbar}\int \! d^3\bR\,|\Omega_e|^2\,n_F(\varepsilon^\bR_{\bk\beta}-\hbar\omega-\mu) \\
\times
A(\bk-\bq,\varepsilon^\bR_{\bk\beta}-\hbar\omega-\mu;\mu_\bR).
\eqname{LDA2}
\end{multline}
The only difference with \eq{LDA1} is that the energy dispersion
of $\beta$ atoms now depends on position according to
\begin{equation}
\varepsilon^\bR_{\bk\beta}=\varepsilon_{\bk\beta}+V_\beta(\bR).
\end{equation}
It is important to stress that the validity of \eq{LDA2} requires
more stringent conditions on the measurement process than a
standard LDA. First, the out-coupled $\beta$ atoms must not
significantly move during the Raman process nor be accelerated.
Second, the trap potential for $\beta$ atoms should be switched
off as soon as possible after the Raman process in order to
minimize the distortion of the $\bk$-space distribution of the
out-coupled atoms induced by the evolution in the trapping
potential.

Equation \eq{LDA2} is the key result of this section and will be
the base of the calculations that we shall present in the
following Sections.

\subsection{Localized probe}

For the sake of simplicity, we restrict our attention to the case
of a pair of Raman beams at an angle $\phi$ with transverse
Gaussian envelopes (Fig.\ref{Raman}),
\begin{eqnarray}
\tilde{\bE}_{1}(\br)&=&\bE_{1}^o\, e^{-z^2/2\sigma^2-y^2/2\sigma^2}\\
\tilde{\bE}_{2}(\br)&=&\bE_{2}^o\, e^{-z^2/2\sigma^2-(x\sin\phi-y\cos\phi)^2/2\sigma^2}.
\end{eqnarray}
The Raman coupling amplitude has therefore a Gaussian shape localized around $\br=0$
\begin{equation}
\Omega_{e}(\br)=\Omega_{e}^o\,e^{-\br^T\bL\br/2\sigma^2}
\end{equation}
with a peak amplitude
\begin{equation}
\Omega_{e}^o=\frac{(\bE_{1}^o \cdot \bd_{\gamma\a})(
\bE^{o*}_{2}\cdot d_{\b\gamma} )} {\Delta}
\end{equation}
and a Gaussian matrix
\begin{equation}
\bL=\left(%
\begin{array}{ccc}
  \sin^2\phi & \sin\phi\cos\phi & 0 \\
  \sin\phi\cos\phi & 1+\cos^2\phi & 0 \\
  0 & 0 & 2 \\
\end{array}%
\right).
\end{equation}
Inserting this formulas into the general expression for the Raman
rate \eq{RamanRateA} and then changing to relative $\rho=\br-\br'$
and center-of-mass variables $\bR=(\br+\br')/2$, one obtains
\begin{multline}
\bR_{\bq}(\bk,\omega)= \frac{2\pi\,|\Omega_e^o|^2}{\hbar}\, \int
\! d^3\bR\, d^3\rho \, e^{-\bR^T \bL
\bR/\sigma^2} \, e^{-\rho^T\bL \rho/4\sigma^2} \\
\times e^{i(\bq-\bk)\rho}\,
n_{F}(\varepsilon_{\bk\beta}-\hbar\omega-\mu)\,
A(\bR,\rho;\varepsilon_{\bk\beta}-\hbar\omega-\mu).
\end{multline}
For the sake of completeness, it is interesting to briefly discuss
the case of a spatially homogeneous system in the absence of any
trapping potential that is probed by tightly focussed Raman beams.
This formula can be further simplified by performing the Gaussian
integrals, which leads to
\begin{multline}
\bR_{\bq}(\bk,\omega)= \frac{(2\pi)^4\,|\Omega_e^o|^2\,
\sigma^6}{\hbar\,\textrm{det}[\bL]}\,
\int \! d^3\bq'\,e^{-\sigma^2\,(\bq-\bq')^T \bL^{-1} (\bq-\bq')}\,\\
\times n_{F}(\varepsilon_{\bk\beta}-\hbar\omega-\mu)
\,A(\bk-\bq',\varepsilon_{\bk\beta}-\hbar\omega-\mu).
\eqname{RamanRateLoc}
\end{multline}
As compared to the result \eq{RamanRateAk} for the spatially
homogeneous geometry, the momentum distribution of the Raman
out-coupled atoms is now smeared out by the convolution with the
Fourier transform of the Raman extraction profile. As expected,
this latter has a width of the order $1/\sigma$.

A faithful image of the spectral and momentum features of the
system is then obtained simply by choosing a value of $\sigma$
much larger than all microscopic length scales of the system. This
condition is compatible with the spatial selection of an almost
homogeneous region as soon as the system is macroscopic, i.e. has
a much larger size than all microscopic length scales.

It is interesting to note that the Gaussian factor in
\eq{RamanRateLoc} tends to a delta-function in the
$\sigma\rightarrow \infty$ limit, which perfectly recovers the
results of Sec.\ref{sec:homogeneous} for the spatially homogeneous
system.  On the other hand, all information on the momentum
distribution is lost in an extremely localized Raman probe of size
$\sigma\ll k_F^{-1}$ which provides momentum-integrated images for
a given $\omega$.

\section{Non-interacting fermions: RF vs. Raman}
\label{sec:nonint}

Once we have established a general theory of Raman spectroscopy,
it is important to validate its performances on some specific
examples whose physics is well-known and under control. In this
Section, we shall start from the simplest case of a
non-interacting degenerate Fermi gas. In particular, we shall
discuss the effect of the trapping potential and we shall
demonstrate the advantages of Raman spectroscopy over RF
spectroscopic in order to obtain spatially resolved information on
the microscopic properties of the system even in the presence of a
trap.

Throughout the whole section we shall use the parameters of the
Boulder experiment~\cite{Jin:RFspectroscopy}: (i) a Fermi energy
$E_F=h\times 9.4\,\textrm{kHz}$, (ii) a Fermi wave vector
$k_F=8.6\,\mu \textrm{m}^{-1}$, a total number of atoms $N=3\times
10^5$, (iv) a total size of the cloud $L=190/k_F$, and (v) a
temperature $T/T_F=0.18$. As the total size of the cloud
$L=190/k_F$ is much larger than the only microscopic length scale
$k_F^{-1}$, the system can be safely considered in the macroscopic
regime where the Local Density Approximation is accurate.

In order the help the reader to extract the significant
information on the dispersion of $\alpha$-state quasi-particles,
all plots from now on (except the left panel of
Fig.\ref{fig:RamanSpectroscopy}) will be given as a function of
$E_s=\varepsilon_{\bk\beta}-\hbar\omega$ and $\bar{\bk}=\bk-\bq$.

\subsection{RF spectroscopy}
\begin{figure}[!ht]
  \includegraphics[width=8.5 cm,clip=true]{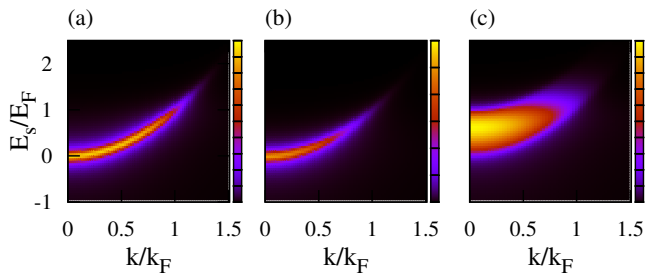}
 \caption{Contour plot of the calculated RF intensity
 for a degenerate gas of non-interacting Fermions at $T/T_F=0.18$.
The color scale indicates the rate of transfer into a state of
momentum $\bk$ of the $\beta$ state as a function of the
(rescaled) RF frequency $E_s=\varepsilon_{\bk\beta}-\omega$. The
left (a) panel refers to a spatially homogeneous system. The (b,c)
panels have been calculated within the Local Density
Approximation: The central (b) panel refers to the case where both
$\alpha$ and $\beta$ atoms experience the same harmonic trap
potential. The right (c) panel refers to the case in which only
the $\alpha$-atoms are trapped. The RF field is spatially uniform
over the whole system.} \label{fig:RFspectroscopy}
\end{figure}

Simulated plots for the RF signal are given in
Fig.\ref{fig:RFspectroscopy} for different trapping
configurations. As usual for RF spectroscopy, the RF field acts
uniformly on the whole atomic cloud and the transferred momentum
is $\bq=0$, which implies $\bar{\bk}=\bk$.

The left panel Fig.\ref{fig:RFspectroscopy}(a) refers to the case
of a spatially homogeneous system in the absence of any trapping
potential. As the dispersion of the $\alpha$ and $\beta$ states of
non-interacting Fermions are exactly parallel, the
momentum-resolved RF signal peaks at  the same value $\omega_{\rm
res}=\varepsilon_{\bk\beta}-\varepsilon_{\bk\alpha}$ independently
of $\bk$. Once we move to the $(\bar{\bk},E_s)$ variables considered
in the plot, the peaks lies on the dispersion
$\varepsilon_{\bk\alpha}$ of the $\alpha$ state. The peak
intensity is independent of $k$ and extends up to the edge of the
Fermi sphere at $k_F$. The abrupt edge at $k_F$ is here smoothened
by the finite temperature value $T/T_F=0.18$.

\begin{figure}[!ht]
  \includegraphics[width=7.cm,clip=true]{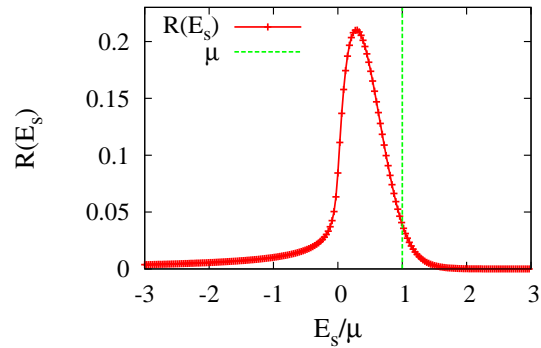}
  \caption{Momentum-integrated RF signal for a trapped gas; as in the central panel of
Fig.\ref{fig:RFspectroscopy}(b), the trap potential is assumed to
act in the same way on the two $\alpha,\beta$ atomic states.
Differently from \eq{total}, the momentum integration is here
performed along lines of fixed $E_s$. The calculated spectrum is
in good agreement with the experimental observation by the Boulder
group (Fig.5b of Ref.~\onlinecite{Jin:RFspectroscopy})}
\label{Jin2}
\end{figure}

The effect of trapping is included in the central and right
panels, Fig.\ref{fig:RFspectroscopy}(b,c). In the central panel,
the $\alpha$ and $\beta$ states are assumed to experience the same
trapping potential as in the Boulder experiment. Under this
condition, the dispersions of the $\alpha$ and $\beta$ states
remain parallel at a fixed distance, so that it is still possible
to extract the $\alpha$ state dispersion following the peak of the
signal intensity in the $(\bar{k},E_s)$ plane. In contrast to the
previous case, the peak intensity is however strongly dependent on
$k$: while the low-momentum states are filled at all positions of
the trap, only the center of the trap contributes in fact to the
high-momentum states close to $k_F$. This implies that the peak
intensity strongly decreases with $k$ and, in particular, it can
be hardly visible in the region around $k_F$. This can be a
serious issue when one is to assess the effect of interactions,
which is generally most pronounced in the vicinity of the Fermi
edge.

Note the qualitative agreement of the calculated signal of
Fig.\ref{fig:RFspectroscopy}(b) with the experimental observations
by the Boulder group shown in Fig.3a of
Ref.~\cite{Jin:RFspectroscopy}. The agreement is even more
striking when we compare the theoretical plot for the
$\bk$-integrated signal shown in Fig.\ref{Jin2} and the
experimental result in Fig. 5b of Ref.~\cite{Jin:RFspectroscopy}.

The effect of the trap is even more dramatic if we consider the
case where only the $\alpha$ state feels the trap potential while
the $\beta$ atoms are untrapped. The distance between the
$\varepsilon^\bR_{\bk\alpha}$ and $\varepsilon^\bR_{\bk\beta}$
dispersions is now strongly dependent on the position $\bR$, which
produces the sizable inhomogeneous broadening that is visible in
Fig.\ref{fig:RFspectroscopy}(c): the lower boundary corresponds to
the contribution of the central region of the trap, while the
upper boundary corresponds to the edges of the cloud. In this
case, it appears to be difficult to extract useful information on
the $\varepsilon_{\bk\alpha}$ dispersion from such a broadened RF
signal.

\subsection{Raman spectroscopy}

\begin{figure}[!ht]
  \includegraphics[width=7.5cm,clip=true]{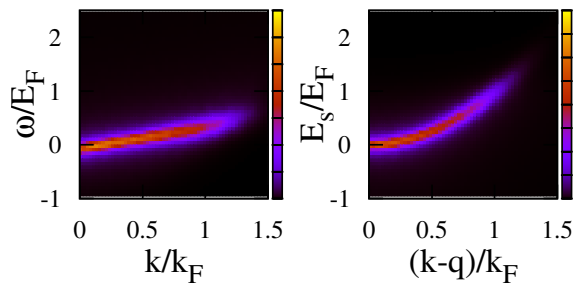}
 \caption{Contour plot of the Raman rate for a trapped
 gas with the same parameters as in the Boulder experiment and a
beam transverse size $\sigma=15/k_F$.
 Left panel: raw data in the $(k,\omega)$ plane. Right panel: Translated data in the
 $(|\bk-\bq|,E_s)$ plane.}\label{fig:RamanSpectroscopy}
\end{figure}

The advantage of the Raman spectroscopy technique over the RF one
is clearly visible in the simulated Raman signal that is plotted
in Fig.\ref{fig:RamanSpectroscopy}. For the parameters of the
Boulder experiment of Ref.~\onlinecite{Jin:RFspectroscopy}, a
transverse size of the Raman beams $\sigma=15/k_F\simeq
1.74\,\mu\textrm{m}$ well satisfies the inequalities $L\gg \sigma
\gg k_F^{-1}$. In particular, we consider the Raman beams to be
focussed onto the central region of the trap and to have an angle
$\phi=\pi/2$ between them. Their wavelength is assumed to be in
the optical range, e.g. $\lambda=0.7\,\mu\textrm{m}$. The momentum
transferred to the atoms during the Raman process is then
$|\bq|=2\sin (\phi/2)/\lambda \approx 2 \mu \textrm{m}^{-1}$, much
smaller than the Fermi momentum $q/k_F\approx 0.2$.

Raw data for the momentum-resolved Raman rate as a function of
$\omega$ and $k$ are plotted in the left panel
Fig.~\ref{fig:RamanSpectroscopy}(a). In order to facilitate
physical understanding of the quasi-particle dispersion, the same
data are plotted in the right panel as a function of the rescaled
variables $E_s=\epsilon_{\bk\beta}-\omega$ and
$\bar{\bk}=\bk-\bq$. Even though we are dealing with a trapped
gas, the observed signal is now very similar to the one of the
homogeneous system shown in Fig.\ref{fig:RFspectroscopy}(a): this
proves the utility of the spatial selectivity of Raman
spectroscopy in view of extracting information on the local
microscopic properties of a trapped gas. The broadening due to the
finite beam size $\sigma=15/k_F$ is almost irrelevant on the scale
of the graphs.

\section{Strongly correlated fermions}
\label{sec:strongly}

To better understand the advantage of Raman
spectroscopy over the RF technique,  we now extend our analysis to
a remarkable example of strongly interacting fermionic system.
Inspired by the Boulder experiment of
Ref.\onlinecite{Jin:RFspectroscopy}, we consider (i) a Fermi gas
in the unitary limit $(k_F a_s)^{-1}=0$ at a temperature close to
the superfluid critical temperature where the physics is dominated
by a pseudogap, and (ii) a molecular gas in the BEC regime $(k_F
a_s)^{-1}>1$. In both cases, we will show how the Raman
spectroscopy is able to provide useful information on the
quasi-particle spectrum of the system.

\subsection{Preformed pairs at unitary limit}

At a temperature just below the superfluid critical temperature
$T/T_c=0.9$, the superfluid fraction and the superfluid order
parameter are still very small. Yet, a sizable pairing gap appears
in the spectral function due to the presence of preformed pairs.
Remarkably, such a pseudo-gap has a width of the order of the
Fermi energy in the critical region and persists even at
temperatures well above the critical point.

\begin{figure}[!ht]
\begin{minipage}[c]{.54\linewidth}
  \includegraphics[height=4.5cm,width=4.8cm,clip=true]{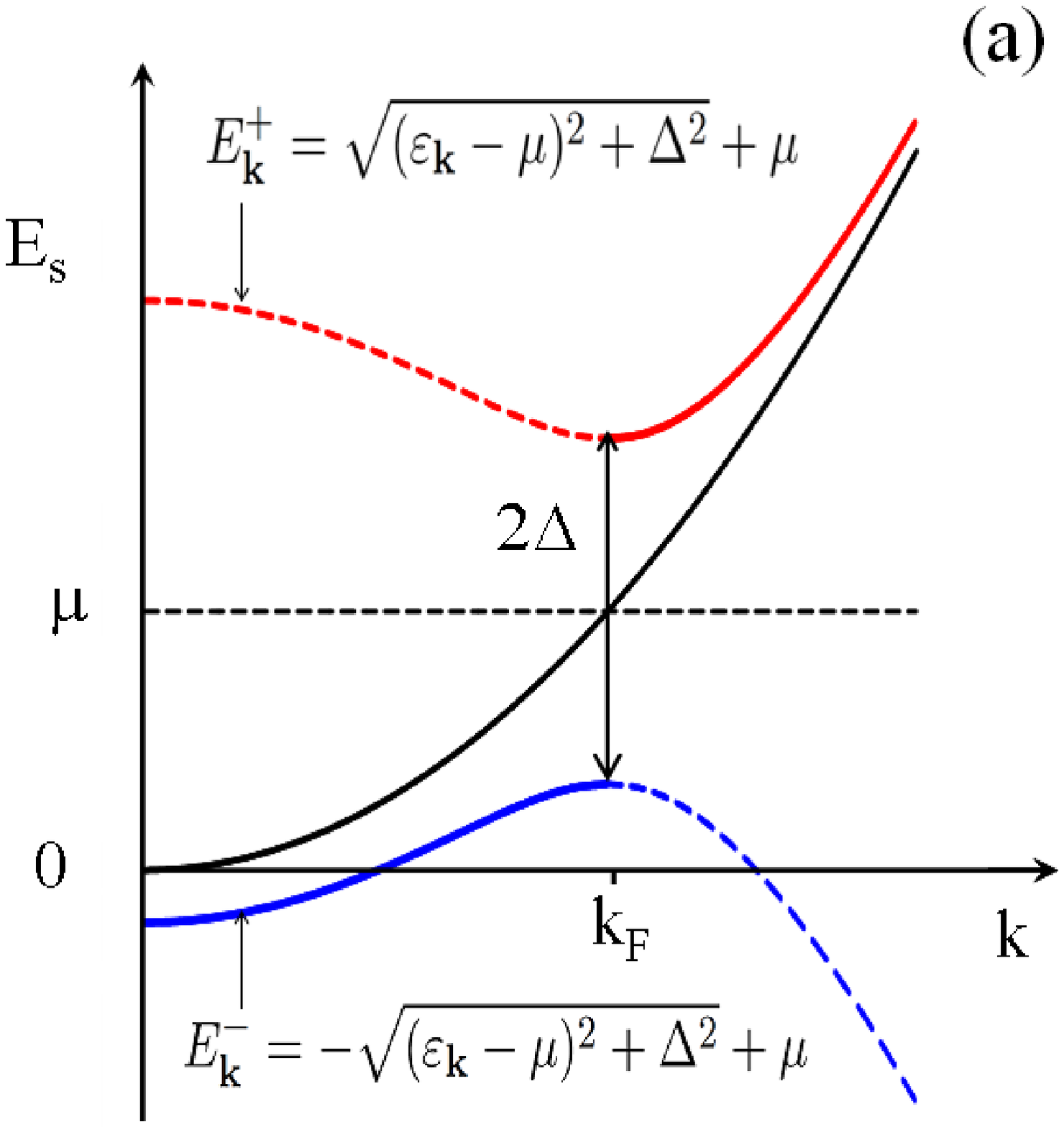}
\end{minipage}\hfill
\begin{minipage}[c]{.44\linewidth}
  \includegraphics[height=6.5cm,clip=true]{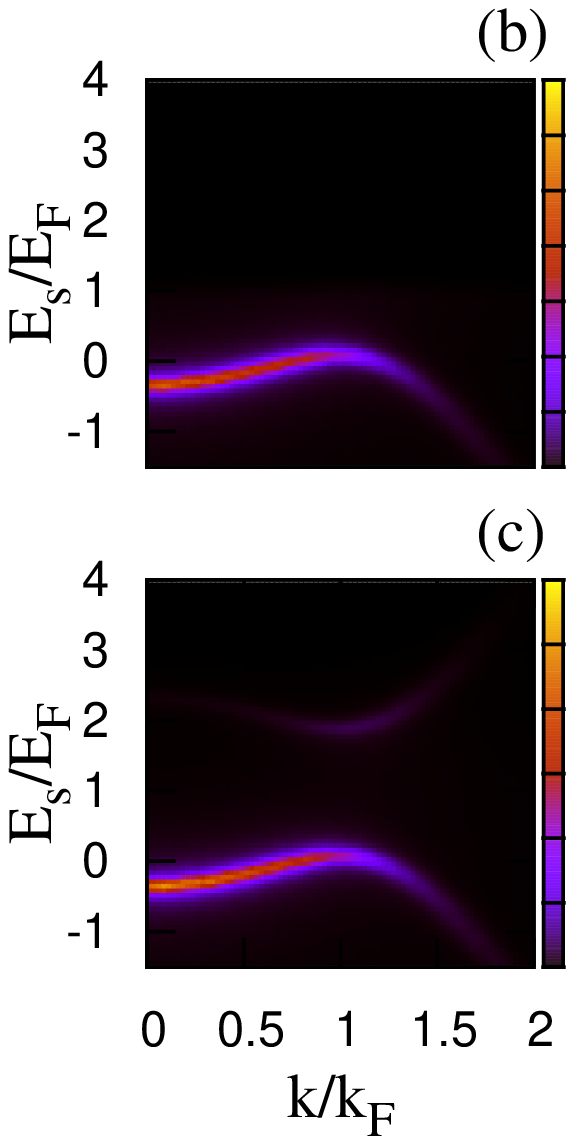}
\end{minipage}\hfill
  \caption{Left (a) panel: Pseudogap model for a Fermi gas in the unitary limit.
  The black line is the free particle dispersion while the blue and red curves are the BCS quasi-particle branches.
  Central (b) and right (c) panels: contour plots of the RF rate for a spatially homogeneous system at
low $T/T_F=0.05 $ and intermediate $T/T_F=0.4$ temperatures.
  }
  \label{UnitaryPreformedPair}
\end{figure}

A reasonable picture of the spectral function in the presence of a
pseudogap is obtained using the standard BCS theory with the
pseudogap $\Delta_{ps}$ replacing the standard superfluid gap
$\Delta_{SF}$,
\begin{equation}
A(\bk,\omega)=u^2_{\bk}\,\delta(\hbar\omega+\mu-E^{+}_{\bk})+v^2_{\bk}\,
\delta(\hbar\omega+\mu-E^{-}_{\bk}) \eqname{BCS_A}
\end{equation}
The quasiparticle energy dispersion $E_\bk^\pm$ has the usual BCS form
\begin{equation}
E^{\pm}_{\bk}=\mu\pm\sqrt{(\varepsilon_{\bk\alpha}-\mu)^2+\Delta^2_{ps}},
\eqname{BCS_qp}
\end{equation}
as well as the Bogoliubov coefficients $u^{2}_{\bk}$ and
$v^{2}_{\bk}$,
\begin{equation}
u^2_\bk,\,v^2_\bk = \frac{1}{2}\left[1\pm \frac{\varepsilon_{\bk\alpha}-\mu}{\sqrt{ (\varepsilon_{\bk\alpha}-\mu)^2+\Delta_{ps}^2 } } \right].
\eqname{uv}
\end{equation}
The RF and Raman signals for a spatially homogeneous Fermi gas at
unitarity are readily obtained by inserting the spectral function
\eq{BCS_A} into the general expression \eq{RamanRateAk}.

At very low temperatures $k_B T/\Delta_{ps}\ll 1$, only the second
term contributes to \eq{BCS_A}: the density of unpaired atoms is
in fact exponentially suppressed by the factor
$\exp(-\Delta_{ps}/k_B T)\ll 1$.  As shown in
Fig.\ref{UnitaryPreformedPair}(b), the RF rate then peaks on a
single branch in the $(k,E_s)$ plane corresponding to the lower
BCS branch at $E_\bk^-$. This branch is strong up to the edge of
the Fermi sphere. The long tail at high momenta is due to the
particle-hole mixing characteristic of BCS theory that is visible
in the expression \eq{uv} of the $v_\bk^2$ Bogoliubov coefficient.

When the temperature gets higher, a second branch appears that
corresponds to the upper branch of the BCS dispersion at
$E_\bk^+$. The weight of this branch rapidly grows with
temperature as $\exp(-\Delta_{ps}/k_B T)\ll 1$. As one can see in
Fig.\ref{UnitaryPreformedPair}(c), the combined effect of the
$u_\bk^2$ coefficient and of the Fermi factor makes the intensity
of this branch to be concentrated in the region close to the
pseudo-gap where most of the unpaired particles are found.

\begin{figure}[!ht]
  \includegraphics[width=7.5cm,clip=true]{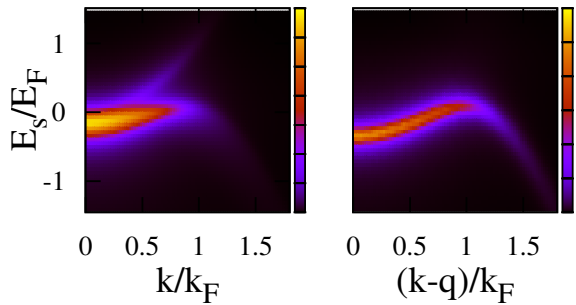}
 \caption{Photo-emission signal for a strongly interacting, trapped Fermi gase at unitarity $(1/k_Fa_s=0)$ at a temperature $T=0.18T_F$. The left panel shows the RF signal. The right
panel shows the Raman signal for a spatially selective process
with $\sigma=15/k_F$. For a comparison of the left panel to
experiment, see Fig.3b of
Ref.\onlinecite{Jin:RFspectroscopy}.}\label{fig:RFPreformedPair}
\end{figure}

\begin{figure}[!ht]
\begin{center}
\includegraphics[width=7cm,clip=true]{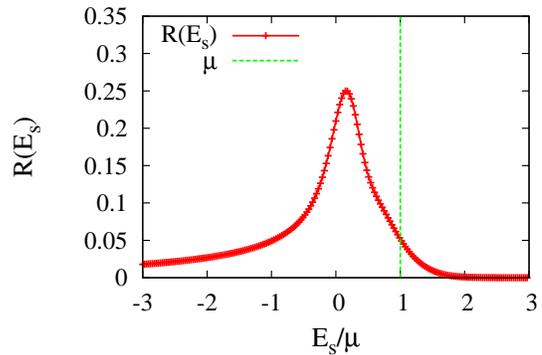}
\caption{Momentum-integrated RF signal for a trapped, strongly
interacting Fermi gas at unitarity. Same system as in the left
panel of Fig.\ref{fig:RFPreformedPair}.  As in Fig.\ref{Jin2},
momentum integration has been performed along lines of constant
$E_s$. For a comparison to experiment, see Fig.5c of
Ref.\onlinecite{Jin:RFspectroscopy}.}\label{Jin4}
\end{center}
\end{figure}

A quantitative comparison to the experimental data shown in Fig.3b
of Ref.~\onlinecite{Jin:RFspectroscopy} requires that we include
in the theoretical model the effect of trapping. This is done in
Fig.\ref{fig:RFPreformedPair}(a): both atomic states $\alpha$ and
$\beta$ are assumed to feel the same trap potential. As we are
working at the unitary limit at temperatures much lower than the
pseudo-gap energy, this latter can be approximated to be
proportional to the local Fermi energy which gives:
\begin{equation}
\Delta_{ps}(\br)=\Delta_{ps}(0)\,\left( 1-\frac{V_{\alpha}(\br)}{\mu} \right).
\end{equation}
The broadening of the line that is apparent at $k=0$ results from
the spatial variation of the minimum of the quasi-particle
dispersion,
$E_{min}=\mu-V_\alpha(\br)-\sqrt{[\mu-V_\alpha(\br)]^2+\Delta^2}$.
The second branch that around $k/k_F\approx 0.8$ emerges from the
Bogoliubov dispersion in the upwards direction is due to the
reduced gap amplitude in the outer part of the cloud and to the
corresponding unpaired atoms. The qualitative agreement of
theoretical data with the experiment is satisfactory. A more
quantitative comparison between theory and experiment (Fig.5c of
Ref.~\onlinecite{Jin:RFspectroscopy}) is successfully made on the
momentum-integrated spectral density shown in Fig.\ref{Jin4}.

The advantage of the Raman spectroscopy technique over the RF one
is apparent in Fig.\ref{fig:RFPreformedPair}(b), where the
simulated signal for a Raman probe localized in the central part
of the trap is plotted. As expected, the inhomogeneous broadening
effects disappear and the signal closely follows the Bogoliubov
branch. Furthermore, the intensity of the branch is everywhere
determined by the Bogoliubov $v_\bk$ coefficient and does not
involve any spatial average. For the parameters considered here
($T/T_F=0.18$, $\Delta\sim E_F$), the intensity of the upper
Bogoliubov branch of unpaired atoms due to the thermal excitations
is relatively small.

\subsection{Tightly bound molecules in BEC limit}

The same BCS theory that we have used in the previous section to
describe the unitary gas can be extended to the BEC side where the
gas is constituted by tightly bound molecules. The spectral
function has the form \eq{BCS_A}, but the gap $\Delta$ and the
chemical potential $\mu$ appearing in the quasi-particle branches
\eq{BCS_qp} have to be calculated by means of the self-consistency
relations of BCS theory.

In the simplest case far in the BEC limit $k_F a_s \ll 1$ at $T\simeq 0$, one has:
\begin{eqnarray}
\mu&=&-\frac{E_b}{2}+\frac{2}{3\pi} E_F (k_F a_s) \\
\Delta&=&\sqrt{\frac{16}{\pi}} \frac{E_F}{\sqrt{k_F a_s}} \\
E_b&=&\frac{\hbar^2}{ma_s^2},
\end{eqnarray}
which corresponds to a pair of quasi-particle branches
\begin{eqnarray}
E_\bk^+&\simeq &\varepsilon_{\bk\alpha}+\frac{12\pi\hbar^2 n a_s }{m}\\
E_\bk^-&\simeq &-E_b-\varepsilon_{\bk\alpha}-\frac{10\pi\hbar^2 n
a_s }{m}.
\end{eqnarray}
The upper branch corresponds to thermally excited unpaired atoms.
As a consequence of the (repulsive) potential of molecules, the
branch is slightly blue-shifted with respect to the free atom
dispersion. The lower branch corresponding to molecules is
separated by the upper one by the binding energy $E_b$ of a
molecule and has an inverted dispersion with negative effective
mass. As suggested in~\cite{Jin:RFspectroscopy}, this peculiar
fact has a transparent physical interpretation: eliminating a
$\alpha$ atom of wavevector $\bk$ means that one has to break a
molecule and eventually leaves the system with an unpaired
$\alpha'$ atom at a momentum $-\bk$. Not only this requires the
binding energy $E_b$ of the molecule, but also the kinetic energy
$\varepsilon_{-\bk\alpha'}$ of the unpaired atom.

\begin{figure}[!ht]
  \includegraphics[width=8.5cm,clip=true]{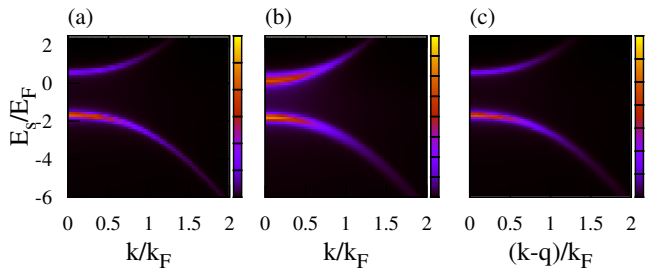}
  \caption{Photo-emission spectroscopy signal for $(k_F a_s)^{-1}=1$ on the BEC side at a temperature $T/T_F=0.5$.
  Left (a) panel: RF signal for a homogeneous system.
  Central (b) panel: RF signal for a trapped system with the parameters of the
experiment of Ref.\onlinecite{Jin:RFspectroscopy}.
  Right (c) panel: Raman signal with a spatially selective probe $\sigma=15/k_F$ focussed onto the cloud center.}
\label{fig:BECMolecule}
\end{figure}

These two branches are visible in the plots of the RF and Raman
signals for a molecular condensate at finite temperature that are
shown in Fig.\ref{fig:BECMolecule}. As usual, the Raman signal
with a spatially selective probe [panel (c)] is able to overcome
the inhomogeneous broadening effects due to trapping that would
otherwise disturb the RF signal [panel (b)] and almost recovers
the RF signal of a homogeneous system [panel (a)].

\section{Inverse Raman spectroscopy}
\label{sec:inverse}

All the discussion so far has been focussed on the case where the
state $\beta$ is initially empty: in this case, momentum-resolved
photo-emission spectroscopy is able to provide detailed
information only on those bands that are initially occupied. This
fact is apparent in the
$n_F(\varepsilon_{\bk\beta}-\hbar\omega-\mu)$ factor multiplying
the spectral function $A$ in the equation for the Raman rate
\eq{RamanRateAk}.

In many cases of current experimental interest, it is however
desirable to have experimental access to the dispersion of empty
bands as well. This is particularly interesting at very low
temperatures, where only the ``negative energy'' states below the
chemical potential are occupied, but much interesting physics is
contained in the ``positive energy'' ones above the chemical
potential.  For instance, in the BCS-like case discussed in
Sec.\ref{sec:strongly}, observation of both $E_\bk^\pm$ branches
would provide unambiguous information on the amplitude of the
superconducting gap.

Information on the empty bands can be obtained if both direct
$\alpha\rightarrow\beta$ and reverse $\beta\rightarrow\alpha$
Raman processes can be induced by the same pair of laser beams.
This is the case if some incoherent population of atoms is already
present in the $\beta$ state at the beginning of the Raman
experiment: the relative occupation of the initial and final
states determines whether the direct or the reverse Raman process
will dominate. The resulting signal results from the difference of
the two photo-emission and photo-absorption processes and is
quantified by the rate of increase/decrease of the population of
the $\bk$ momentum state of the $\beta$ level. Within the same
approximation performed in the previous sections of the paper, the
rate \eq{RamanRateAk} is easily generalized to include also
reverse Raman processes:
\begin{multline}
\frac{dR_{\bq}(\bk,\omega)}{d{V}}=\frac{2\pi}{\hbar}\,|\Omega_e|^2\,
A(\bk-\bq,\varepsilon_{\bk\beta}-\hbar\omega-\mu) \\
\times \left[n_F(\varepsilon_{\bk\beta}-\hbar\omega-\mu)-n_F(\varepsilon_{\bk\beta}-\mu_\beta)\right].
\eqname{RamanRateAkRev}
\end{multline}
The ideal Fermi gas in the $\beta$ state is here assumed to have a
thermal distribution at same temperature $T$ as the gas in the
$\alpha$ state, and a chemical potential $\mu_\beta$.
Generalization of \eq{RamanRateAkRev} to the case of an arbitrary
occupation law is done by simply replacing the occupation factor
$n_F(\varepsilon_{\bk\beta}-\mu_\beta)$. In the
$n_F(\varepsilon_{\bk\beta}-\mu_\beta)=0$ limit,  equation
\eq{RamanRateAkRev} reduces to \eq{RamanRateAk}.

\begin{figure}[!ht]
\includegraphics[width=7.5cm]{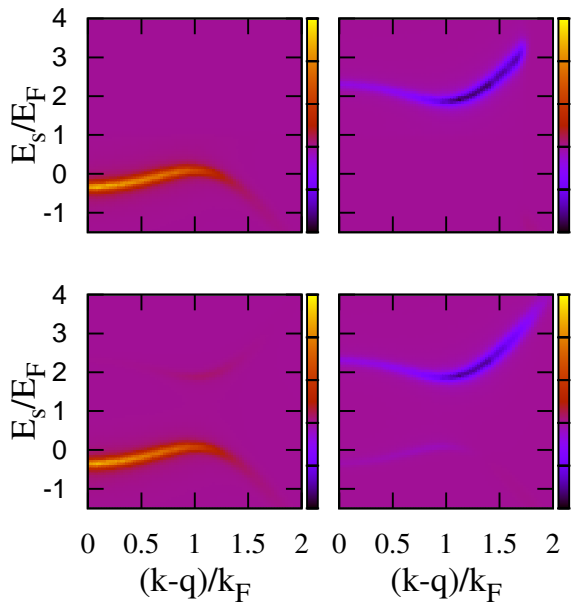}
\caption{Photo-emission spectroscopy with $q=0$ for different
temperatures $T=0.05 T_F$ (upper panels), $T=0.4 T_F$ (lower
panels) and for different initial populations of $\beta$-atoms
$k_{F\beta}=0$ (left panels), $k_{F\beta}=1.73 k_F$ (right
panels). In all panels, the purple background color corresponds to
a vanishing photo-emission and photo-absorption intensity. The
yellow color indicates the occurrence of direct, photo-emission
processes from $\alpha$ into $\beta$; the blue color indicates the
occurrence of reverse, photo-absorption processes from $\beta$
into $\alpha$.}\label{fig:InverseRF}
\end{figure}

In the $\bq=0$ case of RF spectroscopy, probing the upper BCS band
requires that a sizable population is present in the $\beta$ state
up to high momenta. At low temperatures, this requires that the
Fermi momentum $k_{F\beta}$ of the $\beta$ state is larger than
the one $k_F$ of the $\alpha$ state, i.e. that there are initially
more atoms in the $\beta$ state than in the $\alpha$ one. Examples
of combined photo-emission and photo-absorption spectra of the
$\bq=0$ case are shown in Fig.\ref{fig:InverseRF} for the
BCS-pseudogap model of strongly correlated Fermions discussed in
Sec.\ref{sec:strongly} and described by the spectral function
\eq{BCS_A}. Different panels refer to different values of the
initial population of the $\beta$ state (left to right) and to
different temperatures (up to down).

The left panels correspond to an initially empty $\beta$ state: at
$T/T_F=0.05$ (upper panel), only the lower BCS branch at $E_\bk^-$
is visible as a positive, photo-emission signal (yellow). At
higher $T/T_F=0.4$ (lower panel), also the upper BCS branch at
$E_\bk^+$ appears in the spectrum again as a positive signal. The
right panels correspond to a highly degenerate $\beta$ state where
the lowest $\beta$ states have an almost unity filling
$k_{F\beta}=1.73 k_F$. At low temperatures ($T/T_F=0.05$),
photo-emission from the $E_\bk^-$ is strongly suppressed by Pauli
blocking, while the upper band at $E_\bk^+$ clearly appears as a
negative, photo-absorption signal (blue). At higher temperatures,
both bands are visible due to thermal broadening.

\begin{figure}
\includegraphics[width=7.5cm]{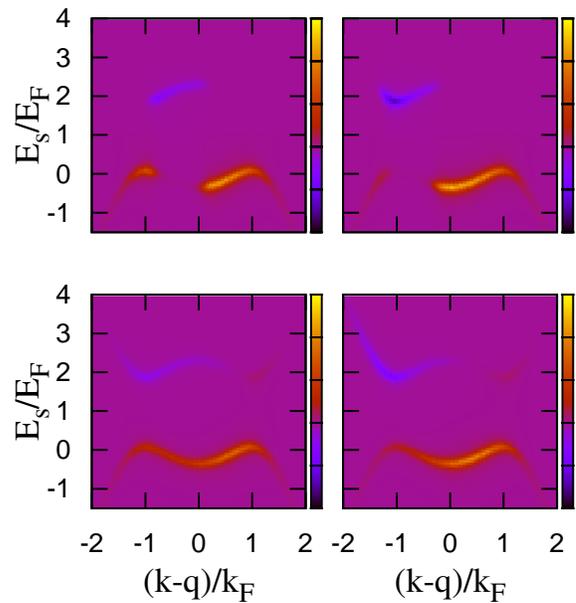}
\caption{Photo-emission spectroscopy for different temperatures
$T=0.05 T_F$ (upper panels), $T=0.4 T_F$ (lower panels) and for
initial population of $\beta$-atoms $k_{F\beta}=0.5 k_F$. The
transfer momentum are: $q=0.5 k_F$ (left panels) and $q=k_F$
(right panels). Same color code as in Fig.\ref{fig:InverseRF}.
}\label{fig:InverseRaman}
\end{figure}

The difficulty of having a high initial density of atoms in
$\beta$ state can be overcome by adopting a Raman scheme with a
transferred wave vector $\bq$ comparable to $k_F$ (see
Fig.\ref{fig:InverseRaman}). In this case, the population in the
almost full lowest $\beta$ states can be effectively transferred
into the empty $\alpha$ states in the most interesting region
$k\simeq k_F$ around the superconducting gap.

Simulated spectra for higher values of the transferred momentum
$q=0.5 k_F,\, k_F$ (left to right) are shown in
Fig.~\ref{fig:InverseRaman}. At low temperature $T/T_F=0.05$
(upper panel), the photo-emission processes from the lower
$E_\bk^-$ branch are suppressed by Pauli blocking in the momentum
range $[-k_{F\beta}-q,k_{F\beta}-q]$. By contrast, the
photo-absorption processes into the upper $E_\bk^+$ branch clearly
appear in this momentum range as a negative signal. At a higher
temperature $T/T_F=0.4$ (lower panel), both processes are
broadened over the whole range of momentum by thermal effects.

The comparison between $q=0.5 k_F$ (left panels) and $q=k_F$
(right panels) shows that the double branch structure is most
visible when the transferred momentum $q$ is closest to the Fermi
momentum $k_F$: in this case, the value of the pseudo-gap
amplitude of the BCS model can be easily extracted from the
distance in frequency between the two features.

\section{Conclusions and perspectives}
\label{sec:conclu}

In conclusion, we have proposed a momentum-resolved Raman
spectroscopy technique that, in analogy to the Angle-Resolved
Photo-Emission Spectroscopy of solid state physics is able to
probe the one-body properties of an atomic gas. The power of this
technique to measure the dispersion of the filled quasi-particle
states has been illustrated on a number of simple systems ranging
from an ideal Fermi gas to a strongly correlated one and has been
validated on existing experimental data.

Several advantages over previous techniques (in particular RF
spectroscopy) have been pointed out: the use of focussed Raman
beams enables one to eliminate the inhomogeneous broadening due to
the trap potential by restricting the optical process to a limited
spatial region. A large value of the transferred momentum can also
help to purify the measured spectra by suppressing the effect of
interactions between the photo-emitted atoms and the rest of the
cloud.

A direct extension of the technique to the case where an
incoherent population is already present in the final state of the
Raman process is finally proposed: Raman processes take place in
both directions and reverse Raman spectroscopy can be used to
obtain information on the empty quasi-particle states lying above
the chemical potential.

Thanks to these remarkable features, we expect that
momentum-resolved Raman spectroscopy will play an important role
in the experimental characterization of the variety of novel
quantum states of matter that can be obtained in systems of
strongly interacting ultracold atoms, e.g. fermionic Mott
insulator states, d-wave superconductors, anti-ferromagnetic
states.

\acknowledgements We are most grateful to C.~Salomon and
J.~Dalibard for collaboration~\cite{DaoPRL} and discussions on
this topic. We also acknowledge useful discussions with
J.-S.~Bernier, C.~Kollath and W.~Zwerger (whom we thank for
informing us of Ref.~\cite{Zwerger_condmat_2009} prior to
submission). I. Carusotto is grateful to the ETH Quantum Optics
group for continuous exchanges. T.-L.~Dao is grateful to the
INFM-BEC center in Trento for hospitality. We acknowledge the
support of the Agence Nationale de la Recherche under contracts
GASCOR and FABIOLA, and of the DARPA-OLE program.

\end{document}